\documentclass[sigconf]{acmart}
\usepackage{amssymb,amsmath}
\usepackage{multirow}
\usepackage{notoccite}

\AtBeginDocument{%
  \providecommand\BibTeX{{%
    \normalfont B\kern-0.5em{\scshape i\kern-0.25em b}\kern-0.8em\TeX}}}

\copyrightyear{2020}
\acmYear{2020}
\setcopyright{acmcopyright}\acmConference[PETRA '20]{The 13th PErvasive Technologies Related to Assistive Environments Conference}{June 30-July 3, 2020}{Corfu, Greece}
\acmBooktitle{The 13th PErvasive Technologies Related to Assistive Environments Conference (PETRA '20), June 30-July 3, 2020, Corfu, Greece}
\acmPrice{15.00}
\acmDOI{10.1145/3389189.3397655}
\acmISBN{978-1-4503-7773-7/20/06}

\begin{document}

\title{Prediction of Epilepsy Development in Traumatic Brain Injury Patients from Diffusion Weighted MRI}

\author{Md Navid Akbar}
\email{makbar@ece.neu.edu}
\orcid{0000-0003-3000-315X}
\affiliation{%
  \institution{Northeastern University}
  \city{Boston}
  \state{MA}
  \country{USA}
  \postcode{02115}
}
\author{Marianna La Rocca}
\email{marianna.larocca@loni.usc.edu}
\affiliation{%
  \institution{University of Southern California}
  \city{Los Angeles}
  \state{CA}
  \country{USA}
}
\author{Rachael Garner}
\email{rachael.garner@loni.usc.edu}
\affiliation{%
  \institution{University of Southern California}
  \city{Los Angeles}
  \state{CA}
  \country{USA}
}
\author{Dominique Duncan}
\email{dduncan@loni.usc.edu}
\affiliation{%
  \institution{University of Southern California}
  \city{Los Angeles}
  \state{CA}
  \country{USA}
}
\author{\text{Deniz Erdo{\u{g}}mu{\c{s}}}}
\email{erdogmus@ece.neu.edu}
\affiliation{%
  \institution{Northeastern University}
  \city{Boston}
  \state{MA}
  \country{USA}
}

\renewcommand{\shortauthors}{Akbar, et al.}

\begin{abstract}
\vspace{-2mm}
Post-traumatic epilepsy (PTE) is a life-long complication of traumatic brain injury (TBI) and is a major public health problem that has an estimated incidence that ranges from $2\%-50\%$, depending on the severity of the TBI. Currently, the patho-mechanism that induces epileptogenesis in TBI patients is unclear, and one of the most challenging goals in the epilepsy community is to predict which TBI patients will develop epilepsy. In this work, we used diffusion weighted imaging (DWI) of 14 TBI patients recruited in the Epilepsy Bioinformatics Study for Antiepileptogenic Therapy (EpiBioS4Rx) to measure and analyze fractional anisotropy (FA), obtained from tract-based spatial statistic (TBSS) analysis.
Then we used these measurements to train two support vector machine (SVM) models to predict which TBI patients have developed epilepsy. Our approach, tested on these 14 patients with a leave-two-out cross-validation, allowed us to obtain an accuracy of $0.857 \pm 0.18$ (with a 95$\%$ level of confidence), demonstrating it to be potentially promising for the early characterization of PTE. 
\end{abstract}

\vspace{-4mm}


\maketitle
\vspace{-2mm}
\section{INTRODUCTION}
\label{sec:intro}
Post-traumatic epilepsy (PTE) is a form of acquired epilepsy that results from a traumatic brain injury (TBI) caused by an external force, for example height falls, motor vehicle accidents, etc. A person with PTE suffers unprovoked and recurrent post-traumatic seizures (PTS) more than one week after the TBI. Effective pharmacological treatments in the prevention or treatment of symptomatic seizures in patients with PTE does not currently exist, therefore the epilepsy community has an urgent need to find reliable biomarkers of PTE \cite{piccenna2017management}. 
Currently, the Epilepsy Bioinformatics Study for Antiepileptogenic Therapy (EpiBioS4Rx) is pioneering this work with the aim to design and perform preclinical trials of antiepileptogenic therapies followed by future planning of clinical trials. Diffusion weighted imaging (DWI), in combination with machine learning techniques, has contributed considerably to the identification of white matter regions affected by neurological conditions in their early phases, especially with the growing dissemination of innovative diffusion tensor imaging (DTI) techniques for tractography~\cite{la2018novel}.

Many DWI studies are based on the measure of quantitative indices, such as fractional anisotropy (FA), radial diffusivity (RD), and mean diffusivities (MD) that characterize  water diffusion in the brain. 
The corpus callosum, superior coronal radiata, cingulate bundle, superior and inferior longitudinal fasciculi, and arcuate fasciculus are the most susceptible white matter tracts after TBI. 
Moreover, structural and functional alterations and disconnections related to seizure development after TBI are found in the thalamus, hippocampus, cingulate gyrus, precentral gyrus, postcentral gyrus, and middle and inferior frontal and temporal gyrus \cite{garner2019imaging, la2019machine, garner2019machine}. 

In this work, we have analyzed a sample of 14 patients from the EpiBioS4Rx cohort: 7 who develop PTE and 7 who have not developed PTE. We apply a tract-based spatial statistic (TBSS) analysis on the diffusion data to obtain FA values to train two support vector machine classifiers (SVMs). 
We then recommend the classifier with the higher accuracy as a promising tool for early detection of PTE.

\begin{figure}[hbt]
    \begin{center}
    \includegraphics[width=0.8\columnwidth]{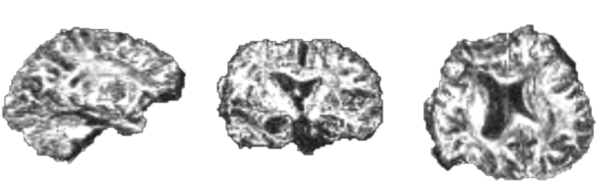}
    \end{center}
    \caption{Cleaned FA map of a sample TBI patient brain.}
    \label{fig_cleaned_FA}
\end{figure}
\vspace{-2mm}

\vspace{-2mm}
\begin{figure}[hbt]
    \begin{center}
    \includegraphics[width=0.7\columnwidth]{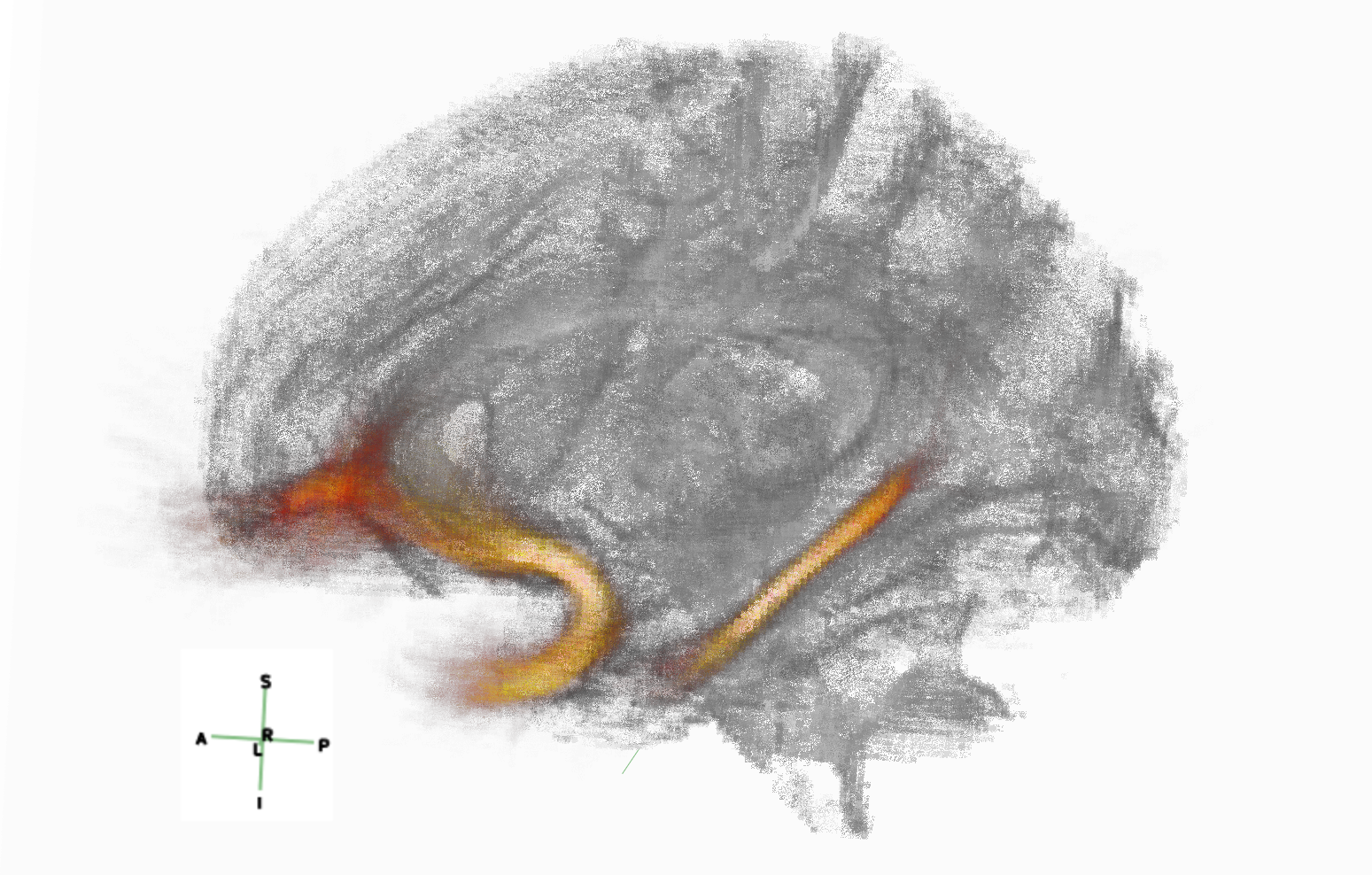}
    \end{center}
    \caption{FA skeleton of a patient registered in the ENIGMA DTI space. The colored tract on the left is the left uncinate fasciculus and the one on the right is the right cingulum.}
    \label{fig_tracts_in_std_space}
\end{figure}

\vspace{-2mm}
\begin{figure}[hbt]
    \begin{center}
    \includegraphics[width=0.8\columnwidth]{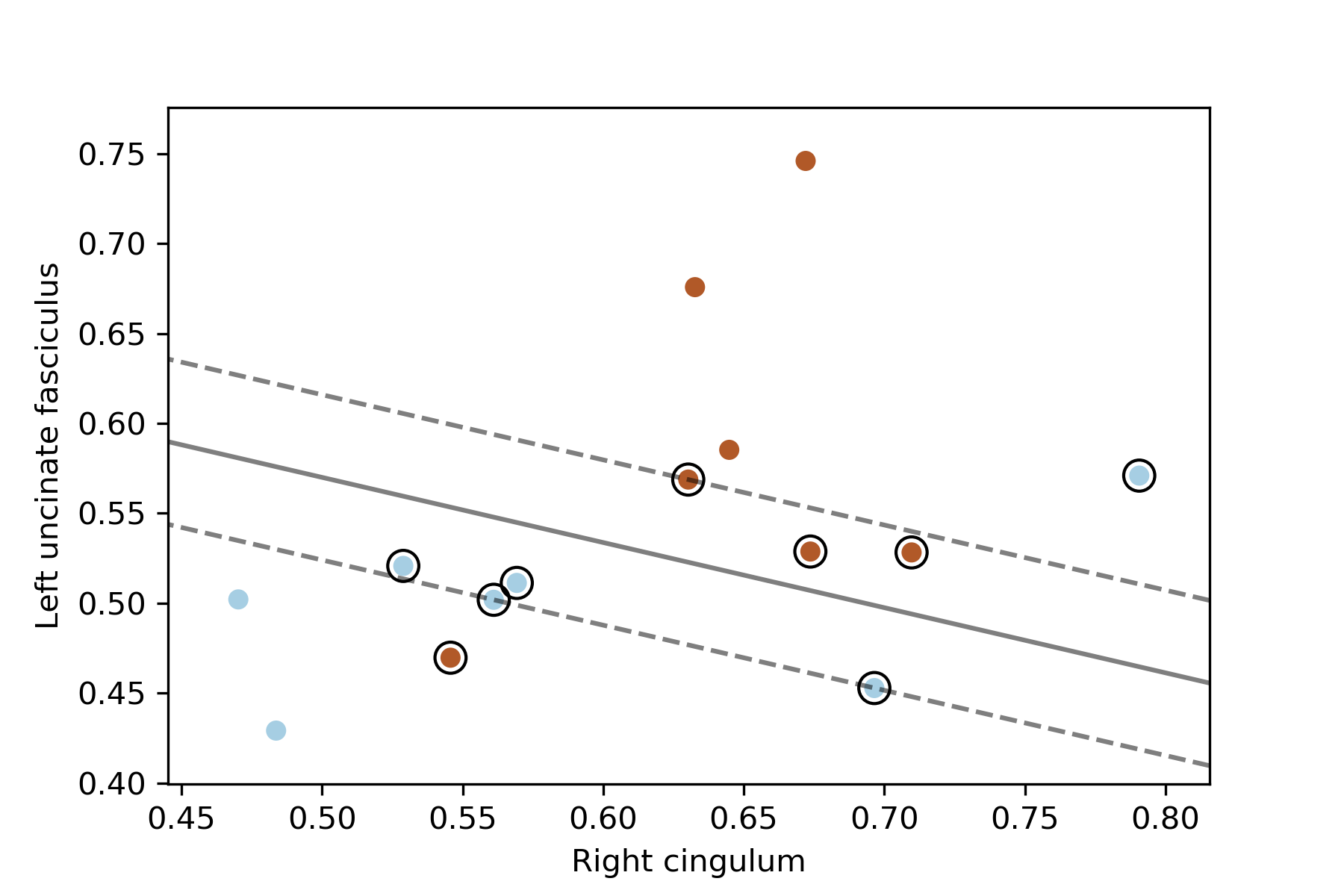}
    \end{center}
    \caption{A linear SVM on the mean FA values along the chosen tracts for all 14 patients. The red dots represent patients who develop PTE, and the blue ones who do not develop PTE.}
    \label{fig_prediction_sample}
\end{figure}
\vspace{-2mm}

\section{METHODOLOGY}
\label{sec:methods}
According to the EpiBioS4Rx protocol, for each enrolled patient, different MRI sequences are acquired within 1-18 post-injury days: T1 MRI, resting state blood oxygenation level dependent (BOLD) functional magnetic resonance imaging (rs-BOLD fMRI), DWI, gradient echo/ susceptibility weighted imaging (GRE/SWI), T2, and  T2-weighted fluid attenuated inversion recovery (FLAIR).

The acquired DWI scans of each patient correspond to multiple diffusion gradient values and directions. 
This DWI data is processed in the FMRIB Software Library (FSL) \cite{smith2006tract}. 
Next, FSL’s Brain Extraction Tool (BET) is applied to create a binary brain mask.
Afterwards, FSL's Diffusion Toolbox (FDT) is used to estimate DTI parameters for FA, MD, etc. image maps.
The FA images are then slightly eroded in FSL to remove brain-edge artifacts and possible outliers.
An example cleaned image is seen in Fig. \ref{fig_cleaned_FA}

In the next phase, the TBSS analysis is performed following the ENIGMA DTI pipeline \cite{stein2012identification}.
The cleaned FA images are first registered to the standard ENIGMA DTI FA space.
Following registration, an FA skeleton is generated for each projected FA map, using the ENIGMA DTI FA standard distance map. 
Along 46 different tracts in each skeleton, obtained from the JHU atlas, the mean FA values are calculated.
In this work, we have chosen the mean FA values of the left uncinate fasciculus and the right cingulum, the two tracts that demonstrate the most distinctive difference between the two groups in our sample, and are also potentially important biomarkers for PTE prediction \cite{garner2019machine}. 
A sample skeleton of a patient's brain, with the chosen tracts, is shown in Fig. \ref{fig_tracts_in_std_space}.

\vspace{-2mm}
\section{Experimental Results}
\label{sec:result}

The scores of the selected features for the two groups are visualized in Fig. \ref{fig_prediction_sample}. 
As evident, the groups appear to be well separable.
Accordingly, we tested two SVMs on the data: one with a linear kernel, and the other with a radial basis function (RBF) kernel, defined by a gamma value of $0.5$. 
The testing is based on a leave-two-out cross-validation, where for each round, the test set contains one patient who will develop at least one seizure, and another who will not develop seizures. 
The analysis results for this sample, for a $95\%$ level of confidence, are tabulated in Table 1. 
As we can observe, the linear SVM actually does a better job in predicting if a patient develops PTE, in terms of both accuracy and F1 score.

\vspace{-2mm}
\begin{table}[htb]
\begin{center}
\caption{Results from SVM Analyses.}
\begin{tabular}{|c|c|c|}
    \hline
    SVM Kernel & Accuracy & F1 Score\\
    \hline
    Linear &  $0.857 \pm 0.18$ & $0.81 \pm 0.28$\\
    \hline
    RBF &  $0.714 \pm 0.20$ & $0.62 \pm 0.33$\\
    \hline 
\end{tabular}
\label{tab_2}
\end{center}
\end{table}
\vspace{-2mm}

\vspace{-1mm}
\section{Conclusion}
\vspace{-1mm}

In this preliminary work, we have demonstrated that a linear SVM trained on FA values derived from a TBSS analysis of diffusion imaging has the potential to detect PTE at an early stage, with a high degree of accuracy and reliability.
A limitation of this work rests in the possible inclusion of false negatives, since here we do not have the follow up data for all patients for the entire two years.


\vspace{-2mm}
\begin{acks}
This project was funded by the National Institute of Neurological Disorders and Stroke (NINDS) of the National Institutes of Health (NIH) under award number R01NS111744.
\end{acks}
\vspace{-2mm}

\bibliographystyle{ACM-Reference-Format}
\bibliography{Refs.bib}

\end{document}